\documentclass[sigconf]{acmart}

\renewcommand\footnotetextcopyrightpermission[1]{}
\usepackage{algorithm}
\usepackage{algpseudocode}
\usepackage{amsmath}
\usepackage{amssymb}  
\usepackage{multirow} 

\usepackage{graphicx}
\usepackage{booktabs}
\usepackage{caption}
\usepackage{subcaption}
\usepackage{hyperref}
\usepackage{float}

\acmConference[Preprint]{arXiv}{arXiv}{2025} 
\acmYear{}
\copyrightyear{}

\begin{document}

\title[TF-DWGNet]{TF-DWGNet: A Directed Weighted Graph Neural Network with Tensor Fusion for Multi-Omics Cancer Subtype Classification}

\author{Tiantian Yang}
\authornote{Corresponding author}
\affiliation{
  \department{Mathematics and Statistical Science}   
  \institution{University of Idaho}
  \city{Moscow}
  \state{Idaho}
  \country{USA}
}
\email{tyang@uidaho.edu}

\author{Zhiqian Chen}
\affiliation{
  \department{Computer Science and Engineering}
  \institution{Mississippi State University}
  \city{Starkville}
  \state{Mississippi} 
  \country{USA}
}      

\renewcommand{\shortauthors}{Yang and Chen}

\begin{abstract}
Integration and analysis of multi-omics data provide valuable insights for improving cancer subtype classification. However, such data are inherently heterogeneous, high-dimensional, and exhibit complex intra- and inter-modality dependencies. Graph neural networks (GNNs) offer a principled framework for modeling these structures, but existing approaches often rely on prior knowledge or predefined similarity networks that produce undirected or unweighted graphs and fail to capture task-specific directionality and interaction strength. Interpretability at both the modality and feature levels also remains limited. To address these challenges, we propose \textbf{TF-DWGNet}, a novel \textbf{G}raph Neural \textbf{Net}work framework that combines tree-based \textbf{D}irected \textbf{W}eighted graph construction with \textbf{T}ensor \textbf{F}usion for multiclass cancer subtype classification. TF-DWGNet introduces two key innovations: (i) a supervised tree-based strategy that constructs directed, weighted graphs tailored to each omics modality, and (ii) a tensor fusion mechanism that captures unimodal, bimodal, and trimodal interactions using low-rank decomposition for computational efficiency. 
Experiments on three real-world cancer datasets demonstrate that TF-DWGNet consistently outperforms state-of-the-art baselines across multiple metrics and statistical tests. In addition, the model provides biologically meaningful insights through modality-level contribution scores and ranked feature importance. These results highlight that TF-DWGNet is an effective and interpretable solution for multi-omics integration in cancer research.
\end{abstract}


\maketitle

\section{Introduction}

Cancer remains a major global health challenge, marked by uncontrolled cell growth, genetic instability, and complex molecular alterations that vary across cancer types and among individual patients. This biological heterogeneity underscores the importance of accurate cancer subtype classification, which is central to precision medicine. Subtype classification enables personalized treatment strategies by leveraging individual molecular profiles, yet reliable prediction remains difficult due to the complexity and multi-layered nature of tumor biology.
Advances in high-throughput technologies now enable simultaneous profiling of multiple omics layers, such as DNA methylation, mRNA expression, and miRNA expression, providing a comprehensive molecular view of tumors \cite{subramanian2020multi, hood2013human}. These modalities offer complementary biological signals that, when jointly analyzed, can reveal disease mechanisms not captured by single-omics analyses. However, multi-omics data integration remains challenging due to high dimensionality, modality heterogeneity, and complex cross-modal interactions \cite{waqas2024multimodal}. Addressing these issues requires computational frameworks that are capable of modeling hierarchical and directional biological processes while preserving model interpretability.

Classical learning approaches such as Random Forest (RF) \cite{breiman2001random}, eXtreme Gradient Boosting (XGBoost) \cite{chen2016xgboost}, and deep feedforward networks (DFNs) \cite{Lecun2015, goodfellow2016deep}, have been widely applied to omics data, but their Euclidean assumptions make it difficult to capture relational structures, handle severe feature dimensionality, or provide task-relevant interpretability \cite{ayman2023review, zaghlool2022review}. 
Geometric deep learning, particularly graph neural networks (GNNs), provides a principled solution by modeling non-Euclidean structures such as biological networks \cite{bronstein2017geometric, zhang2020deep}. GNNs learn node representations by aggregating information from graph neighborhoods \cite{wu2020comprehensive, zhou2020graph}, and architectures such as graph convolutional networks (GCNs) \cite{kipf2017semi}, graph attention networks (GATs) \cite{petar2018graph}, graph isomorphism networks (GINs) \cite{xu2019powerful}, and GraphSAGE \cite{hamilton2017inductive} have demonstrated strong performance in biomedical tasks \cite{zhang2021graph, johnson2024graph}. Meanwhile, multimodal fusion strategies, such as encoder-decoder designs, attention-based methods, and tensor-based models, aim to address heterogeneity across modalities \cite{waqas2024multimodal, zhao2024deep}, although many do not fully capture higher-order interactions.
 
Several GNN-based frameworks have been proposed for multi-omics integration in cancer subtype classification and survival prediction. 
For example, methods such as MOGONET \cite{wang2021mogonet} employ GCNs to process multiple omics modalities and incorporate a cross-omics discovery tensor and a view correlation discovery network for the final prediction. 
Architectures such as MODILM \cite{zhong2023modilm} extend these ideas using graph attention mechanisms to strengthen intra-omics modeling. Other frameworks, including SUPREME \cite{kesimoglu2023supreme} and MOGAT \cite{tanvir2024mogat}, highlight patient similarity networks, GCNs or GATs, and basic concatenation to integrate multiple omics types.  
Additional approaches, such as DeepMoIC \cite{wu2024deepmoic} and MoGCN \cite{li2022mogcn}, leverage autoencoder-based feature extraction and similarity network fusion before applying GNN modules. These approaches differ in terms of how they construct graphs, fuse multimodal information, and ensure interpretability. Despite their contributions, existing frameworks often share several structural limitations: 
(1) \textbf{suboptimal graph construction}: most rely on predefined or correlation-based similarity networks that are undirected, unweighted, and insensitive to task-specific directionality and heterogeneous interaction strengths, both of which are essential for modeling biological regulatory flow;
(2) \textbf{incomplete multi-omics integration}: many approaches use simple concatenation or pairwise modeling and therefore fail to capture higher-order (unimodal, bimodal, trimodal) interactions that may reveal synergistic mechanisms among omics layers;
(3) \textbf{limited interpretability}: interpretability often depends on post hoc analyses, such as feature ablation,  attention inspection, or gradient-based methods, which may not reflect the model's true decision-making process and can be computationally intensive.

To address these challenges, we propose \textbf{TF-DWGNet}, a novel GNN-based framework for interpretable multi-omics integration and cancer subtype classification. Our primary contributions are as follows: (1) \textbf{Supervised directed weighted graph construction}: We leverage XGBoost's tree-splitting structure to construct biologically meaningful, task-specific, and variable-sized directed weighted graphs that encode both directional influence and interaction strength while performing data-driven feature reduction. (2) \textbf{Tensor fusion for unimodal, bimodal, and trimodal interactions}: We implement a low-rank tensor fusion module that efficiently captures higher-order cross-model interactions across all omics types, offering richer representational capacity than concatenation or pairwise fusion. (3) \textbf{Built-in interpretability at both feature and modality levels}: TF-DWGNet inherently produces feature-importance rankings and modality contribution scores without requiring post hoc analyses.
\textbf{End-to-end supervised learning and modality flexibility}: The framework operates end-to-end, from raw omics inputs and labels to final subtype predictions, and can naturally accommodate additional omics modalities without modifying architecture.
To the best of our knowledge, TF-DWGNet is the first framework to jointly integrate supervised directed weighted graph construction, comprehensive low-rank tensor fusion, modality-flexible GNN encoding, and built-in interpretability for multi-omics cancer subtype classification.

\begin{figure*}[h]
  \centering
  \includegraphics[width=\textwidth]{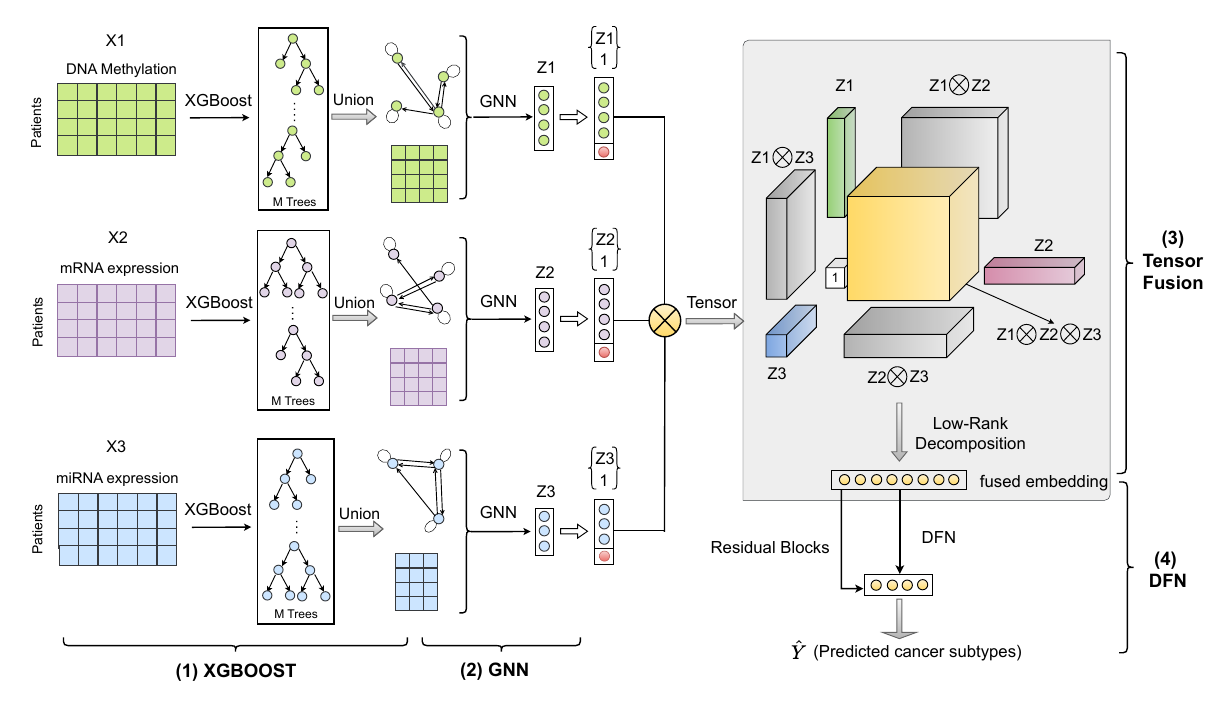}
  \caption{Overview of the TF-DWGNet framework, consisting of four key modules: (i) a XGBoost module for feature selection and supervised construction of directed weighted graphs within each omics modality; (ii) a GNN module that learns unimodal embeddings by jointly encoding graph topology and reduced feature matrices; (iii) a tensor fusion module that models unimodal, bimodal, and trimodal interactions, followed by low-rank CP decomposition; and (iv) a deep residual network for classification. All modules are trained on the training split only; validation and test sets use the learned graphs and parameters without information leakage. TF-DWGNet forms a fully supervised, end-to-end pipeline from raw multi-omics inputs to subtype predictions.
  } \label{fig:TF-DWGNet}
\end{figure*}

\section{Methodology}

\textbf{Problem Definition.}
Let $X \in \mathbb{R}^{n \times p}$ denote the input feature matrix with $n$ samples and $p$ features. The corresponding feature graph is $G(V, E)$, where $V$ is the set of nodes and $E \subseteq V\times V$ is the set of edges. Its adjacency matrix is $A \in \mathbb{R}^{p \times p}$, where $A_{ij}$ encodes the presence (or strength) of an edge from node $i$ to node $j$. We augment the graph with self-loops, giving $\tilde{A} = A + I$, where $I$ is the identity matrix. 
Degree matrix $\tilde{D}$ is diagonal, where each entry is the row sum of $\tilde{A}$.  The normalized adjacency matrix is $\tilde{D}^{-1/2} \tilde{A} \tilde{D}^{-1/2}$.
We consider three omics datasets: $X_1 \in \mathbb{R}^{n \times p_1}$, $X_2 \in \mathbb{R}^{n \times p_2}$, and $X_3 \in \mathbb{R}^{n \times p_3}$, along with multiclass subtype labels $Y \in \mathbb{R}^{n \times 1}$ across $C$ classes, with predictions $\hat{Y} \in \mathcal{Y}$. Given the dataset $\mathcal{D} = \{\{X_i\}_{i=1}^{3}, Y\}$, the objective is to construct the corresponding directed weighted graphs $\mathcal{G} = \{G_i\}_{i=1}^{3}$ and learn a predictive function mapping to $\mathcal{Y}$. Formally, we aim to learn 
$$f_1: \mathcal{D} \rightarrow \mathcal{G} \quad \text{and} \quad f_2: \mathcal{G} \rightarrow \mathcal{Y}$$ 
minimizing the discrepancy between the true labels $Y$ and predicted labels $\hat{Y}$. 

\noindent \textbf{Framework Overview.} Figure~\ref{fig:TF-DWGNet} illustrates the proposed \textbf{TF-DWGNet} framework (\textbf{T}ensor \textbf{F}usion on \textbf{D}irected \textbf{W}eighted \textbf{G}raph \textbf{Net}work),
which integrates four coordinated modules for effective multimodal representation learning: (1) an \textbf{XGBoost module} independently processes each omics modality, performs feature selection, and constructs task-specific directed weighted graphs that capture feature importance and directional relationships; (2) a \textbf{GNN module} learns latent unimodal embeddings from each modality by jointly encoding the reduced feature matrix and its corresponding directed weighted graph; 
(3) a \textbf{Tensor Fusion module} models all unimodal, bimodal, and trimodal interactions across modalities, followed by low-rank Canonical Polyadic (CP) decomposition for computational efficiency; and (4) a \textbf{Deep Residual Network module} takes the compact fused embedding as input and performs multiclass classification using a residual architecture. 
Algorithm~\ref{alg:TF-DWGNet} summarizes the workflow. Overall, TF-DWGNet forms a fully supervised, end-to-end pipeline that takes raw multi-omics inputs and subtype labels as inputs, constructs supervised graphs, learns modality-specific embeddings, performs tensor fusion, and outputs subtype predictions under a unified training objective. 

\begin{algorithm}
\caption{TF-DWGNet} \label{alg:TF-DWGNet}
\begin{algorithmic}[1] 
\Require Omics datasets: $X_1 \in \mathbb{R}^{n \times p_1}$, $X_2 \in \mathbb{R}^{n \times p_2}$, $X_3 \in \mathbb{R}^{n \times p_3}$; multiclass labels $Y \in \mathbb{R}^{n \times 1}$
\Ensure Predicted subtype labels $\hat{Y}$
\State Train XGBoost classifiers on each omics $X_1$, $X_2$, and $X_3$ using $Y$ as the target
\State Select informative features $\Rightarrow$ obtain reduced datasets $X_1^*$, $X_2^*$, $X_3^*$
\State Construct directed, weighted graphs: $G_1^*$, $G_2^*$, $G_3^*$ from tree-split dependencies
\For{each $(X_i^*, G_i^*)$ for $i = 1, 2, 3$}
    \State Compute normalized adjacency matrix with self-loops: $\tilde{D}^{-1/2}\tilde{A}\tilde{D}^{-1/2}$
    \State Feed ($X_i^*, \tilde{D}^{-1/2}\tilde{A}\tilde{D}^{-1/2}$) into a GNN to obtain unimodal embedding $Z_i$
\EndFor  
\State Augment embeddings: $Z_i^*=\begin{bmatrix} Z_i \\ 1 \end{bmatrix}$ for $i=1,2,3$
\State Construct tensor: $\mathcal{T} = Z_1^* \otimes Z_2^* \otimes Z_3^*$ (unimodal, bimodal, trimodal interactions)
\State Apply CP Decomposition:  $\mathcal{T} \approx \sum_{r=1}^R \lambda_r \mathbf{p}_{1, r} \otimes \mathbf{p}_{2, r} \otimes \mathbf{p}_{3, r}$
\State Obtain fused low-rank embedding in $\mathbb{R}^R$
\State Feed fused embedding into a DFN with residual blocks
\State Compute softmax output to obtain $\hat{Y}$
\State \Return Predicted labels $\hat{Y}$.
\end{algorithmic}
\end{algorithm}

\noindent \textbf{Directed Weighted Graph Construction.} 
We leverage XGBoost's sequential tree-building mechanism to capture task-specific feature dependencies and directional interaction patterns in multi-omics data. Unlike random forests, where trees are independent, the boosting structure of XGBoost allows early splits to influence later ones, making it particularly effective at identifying features that serve as upstream stratifiers versus downstream refiners in the decision-making process.
For each omics modality, an ensemble of $M$ boosted trees is trained. Each tree is viewed as a directed graph $G^m = (V^m, E^m)$, where a directed edge $a \rightarrow b$ indicates that feature $a$ appears before $b$ along the same root-to-leaf path. These edges do not represent causal biomarker regulation. Instead, they encode task-specific conditional dependencies: feature $a$ partitions patients into subgroups for which feature $b$ provides additional discriminative power.
Aggregating the feature co-occurrence across all $M$ trees yields a single directed weighted graph $G^*$ as follows: 
$$G^*(V^*, E^*) = G^*(\bigcup_{m=1}^{M} V^{m}, \bigcup_{m=1}^{M} E^{m}),$$ 
where $V^*=\bigcup_{m=1}^{M} V^{m}$ contains all the selected features across $M$ trees and $E^*=\bigcup_{m=1}^{M} E^{m}$ represents directional dependencies, with edge weights equal to their frequencies across trees. Denoting $p^*$ as the number of uniquely selected features, we obtain $|V^*| = p^*$. 
This procedure is applied independently to each omics modality, resulting in three directed weighted graphs: $G_1^* = (V_1^*, E_1^*)$, $G_2^* = (V_2^*, E_2^*)$, and $G_3^* = (V_3^*, E_3^*)$, with corresponding reduced feature matrices $X_1^* \in \mathbb{R}^{n \times p_1^*}$, $X_2^* \in \mathbb{R}^{n \times p_2^*}$, and $X_3^* \in \mathbb{R}^{n \times p_3^*}$, where $p_i^* < p_i$ and the reduced dimensions may differ across modalities. We refer to these XGBoost-selected subsets as \textit{reduced features}. These graphs encode both feature importance and directional structure, providing informative, task-conditioned priors for downstream modeling. \\
\noindent \textbf{Graph Neural Network.} Each modality is then processed using a modified Graph-Embedded Deep Feedforward Networks (GEDFN) \cite{kong2018graph}, which incorporates the adjacency structure by restricting connections between the input and the first hidden layer. 
Let $X^* \in \mathbb{R}^{n \times p^*}$ be the feature-reduced input and $\tilde{A} \in \mathbb{R}^{p^* \times p^*}$ be the directed weighted adjacency matrix with self-loops. 
The first hidden layer is computed as follows:
$$H_1 = \text{ReLU}\left(X^*\left(W\odot (\tilde{D}^{-1/2}\tilde{A}\tilde{D}^{-1/2})\right) + b\right),$$ 
where $W$ and $b$ denote trainable weights and biases, respectively; $\odot$ denotes element-wise (Hadamard) multiplication; and $\tilde{D}$ is the degree matrix. 
Normalization with $\tilde{D}$ prevents the highly connected nodes from dominating the representation.
By embedding directed, weighted feature graphs directly into the first layer, it captures topological dependencies while preserving architectural sparsity, which is an important consideration for modeling high-dimensional multi-omics data.

\noindent  \textbf{Tensor Fusion and Decomposition.} 
Given the unimodal latent representations obtained from GNNs 
$$Z_1 \in \mathbb{R}^{n \times p_1^*}, Z_2 \in \mathbb{R}^{n \times p_2^*}, Z_3 \in \mathbb{R}^{n \times p_3^*}$$ 
for $n$ samples, we employ a \textbf{Tensor Fusion Network (TFN)} \cite{zadeh2017tensor} to model both intra- and inter-modal interactions across the three omics modalities. To enable the unimodal, bimodal, and trimodal interaction terms, each latent representation is augmented with a constant feature: 
$$ Z_1^* = \begin{bmatrix} Z_1 \\ \mathbf{1} \end{bmatrix},  Z_2^* = \begin{bmatrix} Z_2 \\ \mathbf{1} \end{bmatrix}, Z_3^* = \begin{bmatrix} Z_3 \\ \mathbf{1} \end{bmatrix}, $$
where $Z_i^* \in \mathbb{R}^{n \times (p_i^*+1)}$, and the appended 1s facilitate the interaction terms involving fewer than three modalities. 
For each sample, the TFN constructs a third-order tensor using the outer product:
$$\mathcal{T} 
=
\begin{bmatrix} 
Z_1 \\ 
1 
\end{bmatrix}  \otimes 
\begin{bmatrix} 
Z_2 \\ 
1 
\end{bmatrix} 
\otimes 
\begin{bmatrix} 
Z_3 \\ 
1 
\end{bmatrix} \in \mathbb{R}^{(p_1^*+1)\times(p_2^*+1)\times(p_3^*+1)}.
$$
This tensor encodes a complete set of modality interactions:
\begin{itemize}
    \item \textit{\textbf{Unimodal}} terms ($Z_1$, $Z_2$, and $Z_3$):  elements involving only one modality
    \item \textit{\textbf{Bimodal}}  terms ($Z_1 \otimes Z_2$, $Z_1 \otimes Z_3$, and $Z_2 \otimes Z_3$): pairwise multiplicative interactions
    \item \textit{\textbf{Trimodal}}  term ($Z_1 \otimes Z_2 \otimes Z_3$): three-way interactions across all modalities
\end{itemize}

These correspond to the seven structured regions illustrated in the tensor fusion module shown in Figure~\ref{fig:TF-DWGNet}. 
\noindent  While expressive, this tensor scales as
$O\left(\prod_{i=1}^{3} (p_i^*+1)\right)$, which becomes computationally intractable as the latent dimensions increase. 
To address this challenge, we apply a \textbf{Canonical Polyadic (CP) decomposition} (also called CANDECOMP/PARAFAC), \cite{hitchcock1927expression, carroll1970analysis, harshman1970foundations}, 
which approximates the third-order tensor $\mathcal{T}$ by the sum of $R$ rank-one tensors: 
$$\mathcal{T} \approx \sum_{r=1}^R \lambda_r \mathbf{p}_{1, r} \otimes \mathbf{p}_{2, r} \otimes \mathbf{p}_{3, r},$$ 
where $R$ is a positive integer, $\lambda_r  \in \mathbb{R}$, and $\mathbf{p}_{i,r} \in \mathbb{R}^{p_i^* + 1}$ for modalities $i = 1, 2, 3$. 
This formulation generalizes the matrix singular value decomposition (SVD) to higher-order tensors \cite{kolda2009tensor} and reduces the computational complexity to $O\left(R \cdot \sum_{i=1}^3 (p_i^*+1)\right)$. 
\begin{definition}[Rank-One Tensors]
An $N$-way tensor $\mathcal{X} \in \mathbb{R}^{d_1 \times d_2 \times \dots \times d_N}$ is rank-one if it can be written as the outer product of $N$ vectors:
$$\mathcal{X} = \mathbf{a}_{1} \otimes \mathbf{a}_{2} \otimes \dots \otimes \mathbf{a}_{N}.$$
\end{definition}
\begin{definition}[Tensor Rank]
The rank of a tensor $rank(\mathcal{T})$ is the smallest number of rank-one tensors that sum to $\mathcal{T}$. The maximum rank is defined as the largest attainable rank. 
\end{definition} 
\begin{proposition}
For $\mathcal{T} \in \mathbb{R}^{d_1 \times d_2 \times d_3}$, the maximum tensor rank satisfies \cite{kolda2009tensor, kruskal1989rank}: $$rank(\mathcal{T}) \leq min\{d_1d_2, d_1d_3, d_2d_3\}. $$ 
\end{proposition}
Determining the exact tensor rank $R$ is NP-hard in general \cite{haastad1989tensor}, so $R$ is selected by fitting various rank-$R$ CP models and evaluating the performance empirically, consistent with prior tensor literature \cite{carroll1970analysis, harshman1970foundations}. 

\begin{table*}[h]
\caption{Summary of BRCA, UCEC, and KIPAN datasets, including total sample size ($n$), number of classes ($C$), samples per class ($n_c$), $c=1,\dots, C$, original feature counts, and preprocessed feature dimensions for each omics modality. BRCA = Breast Carcinoma; UCEC = Uterine Corpus Endometrial Carcinoma; KIPAN = TCGA Pan-Kidney cohort.
} 
\label{tab:datainfo}
\centering
\setlength{\tabcolsep}{1mm}
\begin{tabular}{@{}llllrr@{}}
\toprule
\textbf{Dataset} & \textbf{$n$} & $C$& \textbf{Samples per class} & \textbf{Original Features} & \textbf{Preprocessed Features} \\
& & &  & (meth:mRNA:miRNA)  & (meth:mRNA:miRNA) \\
\midrule
BRCA & 875 & 5 & Normal-like (115), Basal-like (131), HER2-enriched (46),   & 20,531:20,106:503 & 1,000:1,000:503\\
& &  &   Luminal A (436), Luminal B (147)  \\
UCEC & 430 & 3 & EEA (311), SEA (98), MSEAC (21) & 20,118:20,531:554 & 2,000:2,000:554 \\ 
KIPAN & 707 & 3 & KICH (66), KIPC (343), KIRP (298) & 20,531:20,111:472 & 2,000:2,000:472 \\ 
\bottomrule
\end{tabular}%
\end{table*}

\begin{figure*}[h]
  \centering
  \includegraphics[width=0.9\textwidth]{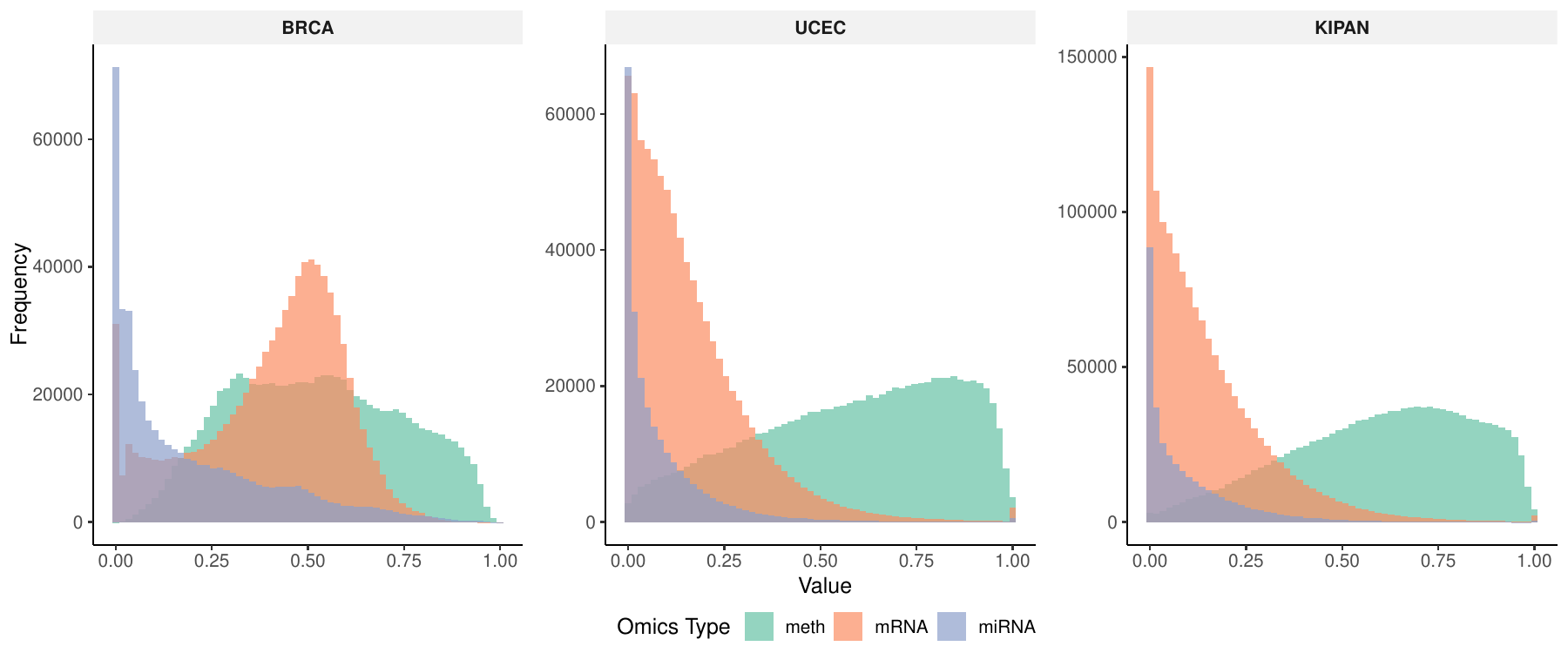}
  \caption{Histograms of DNA methylation, mRNA, and miRNA feature values in the preprocessed BRCA, UCEC, and KIPAN datasets. All features are scaled to the range [0, 1]. The heterogeneous distributions across modalities and datasets highlight the distinct biological characteristics captured by each omics type.} \label{fig:omicsHistogram}
\end{figure*}

\noindent \textbf{Low-Rank Fusion With Modality-Specific Projections.} To support parameter sharing across samples and maintain computational efficiency, we introduce \textit{\textbf{modality-specific projection matrices}}, as follows:
$$\mathbf{P}_1 \in \mathbb{R}^{R\times(p_1^* + 1)}, \mathbf{P}_2 \in \mathbb{R}^{R\times(p_2^* + 1)}, \mathbf{P}_3 \in \mathbb{R}^{R\times(p_3^* + 1)},$$ 
where $\mathbf{P}_i = [\mathbf{p}_{i,1}, \mathbf{p}_{i,2}, \dots, \mathbf{p}_{i,R}]$.  This yields the compact CP-based fusion representation:
\begin{eqnarray*}
\mathcal{T} &\approx& [\![ \mathbf{\lambda}, \mathbf{P}_1 Z_{1}^*, \mathbf{P}_2 Z_{2}^*, \mathbf{P}_3 Z_{3}^* ]\!] \\
&=& \sum_{r=1}^R \lambda_r \cdot (\mathbf{P}_1 Z_{1}^*)_r \cdot(\mathbf{P}_2 Z_{2}^*)_r \cdot (\mathbf{P}_3 Z_{3}^*)_r,
\end{eqnarray*}
where $[\![\cdot, \cdot, \cdot]\!]$ denotes the CP operator. Each term $(\mathbf{P}_i Z_i^*) \in \mathbb{R}^R$ projects the augmented latent vector into rank-$R$ space. The element-wise product across modalities yields a fused tensor representation of $\mathbb{R}^R$ for each sample.  

\noindent \textbf{Novelty.}
Compared to conventional fusion approaches that rely on simple concatenation, full tensor expansion or low-capacity bilinear interactions, our tensor fusion module:
\begin{itemize}
\item models high-order cross-modality interactions through low-rank CP decomposition
\item retains modality-specific projections that capture shared and unique biological structure
\item avoids the computational burden of full tensor expansion
\end{itemize}

\noindent \textbf{Residual Classification Network.} The low-rank fused embedding produced by CP decomposition is then passed to a residual multilayer perceptron \cite{chen2020deep} for final subtype prediction. Each residual block is implemented as either an \textbf{\textit{identity block}} or a \textbf{\textit{projection block}}.
In an identity block, the input and output dimensions match,  allowing the original input to be added directly to the transformed output before applying the activation function. This yields a skip connection of the form:  
$$\sigma(\sigma(\sigma (XW_1)W_2)W_3 + X),$$ 
where $X$ denotes the block input, $W_i$ is the learnable weight matrix, and $\sigma$ is the activation function (bias terms are omitted for simplicity).
When the input and output dimensions differ, a projection block linearly transforms the input to match the hidden width before addition. 
In our architecture, the hidden width is fixed across all layers; therefore, only the first block requires projection, and all subsequent blocks are identity blocks. This design stabilizes the optimization and enables deeper and more expressive feature transformations.

\noindent \textbf{Feature-Level and Modality-Level Interpretability.} 
To provide biologically meaningful interpretability at both the input feature (biomarker) and modality (omics) levels, we adapt and extend the connection weights approach in \cite{olden2002illuminating}. \\
\textbf{\textit{(1) Feature-level importance:}}
For each omics modality, feature importance is derived from the connection weights between the input layer and the first hidden layer of the corresponding GNN. Since the feature graphs are directed, a feature can participate either as a \textit{\textbf{source}} or a \textit{\textbf{target}} node. For example, in edge $a \rightarrow b$, feature $a$ contributes as a source, while in $c \rightarrow a$, it contributes as a target. For feature $j$ in modality $i$, its importance score is defined as:
$$
IF_{j}^{i} = \underbrace{\sum_{u=1}^{p_{i}^*} \left|W_{ju}^{i}I(\tilde{A}_{ju}^{i}\neq 0)\right|}_{\text{source importance} \; IF_{i}^{j, \text{source}}} 
           + \underbrace{\sum_{u=1}^{p_{i}^*} \left|W_{uj}^{i}I(\tilde{A}_{uj}^{i}\neq 0)\right|}_{\text{target importance} \; IF_{i}^{j, \text{target}}},
$$
where $W^{i}$ is the associated weight matrix, and $\tilde{A}^{i}$ is the directed weighted adjacency matrix (with self-loops) for omics modality $i \in 1, 2, 3$. The total importance of feature $j$ is given by: 
$$IF_{j} = \sum_{i=1}^{3}IF_{j}^{i} = IF_{j}^{1} + IF_{j}^{2} + IF_{j}^{3}.$$ \\
\textbf{\textit{(2) Modality-level importance:}} To quantify the relative importance of each omics source, we aggregate the feature-level scores within each modality. Specifically, the model dynamically estimates the relative importance of omics $i$ by:
$$
RIO_{i} = \frac{\sum_{j=1}^{p_i^*}IF_{j}^{i}}{\sum_{i=1}^3\sum_{j=1}^{p_i^*}IF_{j}^{i}},
$$
where $RIO_i \in [0, 1]$ and $\sum_{i=1}^3 RIO_i = 1$. 
This provides a normalized measure of how strongly each omics modality contributes to the final classification, enabling transparent and biologically interpretable insights without requiring additional post hoc analysis or ablations. 

\noindent \textbf{Technical Details.}
XGBoost is implemented using the \texttt{xgboost} Python library, and all GNN and DFN components are trained in \texttt{Tensorflow}. Models are optimized via mini-batch training using the Adam optimizer, with ReLU applied as the activation function throughout and softmax in the output layer for multiclass classification. Training minimizes the categorical cross-entropy loss combined with dynamic $L_2$ regularization applied to all trainable weights from the three GNN branches and the DFN with residual blocks: 
\begin{eqnarray*}
\mathcal{L}_{\text{total}} &=& \underbrace{-\frac{1}{n}\sum_{i=1}^n \sum_{c=1}^C y_{i,c}\ln(\hat{y}_{i,c})}_{\text{Cross-entropy loss}} \\
&+& \lambda \bigg(
\underbrace{\sum_{i=1}^3\sum_{w_i\in \Theta_{G_1}} w_i^2}_{\text{GNN weights}} + \underbrace{\sum_{w\in \Theta_{\text{DFN}}} w^2}_{\text{DFN weights}}\bigg),
\end{eqnarray*}
where 
$y_{i,c}$ is the true label; $\hat{y}_{i,c}$ is the predicted probability; and $\lambda$ is the regularization strength.
We tune the number of estimators for XGBoost and optimize the architectural and training hyperparameters, including depth, hidden width, learning rate, batch size, and maximum epochs, for both the GNN and DFN components. Dropout, early stopping, and batch normalization are applied to mitigate overfitting. Model evaluation is performed over 20 independent stratified training-validation-test splits (60\%:20\%:20\%), ensuring reproducibility through fixed random seeds. In addition, we tune the tensor fusion rank $R$ and the number of residual blocks in the DFN. 
Unless otherwise specified, all components of TF-DWGNet fit exclusively on the training split. For each random split, XGBoost is trained using only training samples, and directed weighted graphs are constructed from tree-splitting paths on this training data. The validation and test samples are then passed through the trained trees to obtain their reduced feature representations, while the graph structure remains fixed within that split. This avoids any leakage of test information into the graph construction stage. All the experiments are executed on the Falcon supercomputer \cite{falcon2022}, with each job requesting four CPUs and 20 GB of memory.

\section{Real Data Experiments}

\textbf{Datasets Overview.}
We evaluate TF-DWGNet on three widely used multi-omics cancer subtype datasets: BRCA, UCEC, and KIPAN. The processed BRCA dataset is obtained from \cite{wang2021mogonet}, while the original UCEC and KIPAN datasets are downloaded from The Cancer Genome Atlas (TCGA) via the Broad GDAC Firehose 
and preprocessed following the procedure in \cite{Lu2023MultiomicsSubtypes}. \textbf{BR}east \textbf{CA}rcinoma (BRCA) is used for PAM50 subtype classification with five molecular subtypes. \textbf{U}terine \textbf{C}orpus \textbf{E}ndometrial \textbf{C}arcinoma (UCEC) contains three clinically relevant subgroups. KIPAN, the TCGA \textbf{PAN}-\textbf{KI}dney cohort, includes three kidney cancer subtypes.   
Table \ref{tab:datainfo} summarizes the sample sizes, number of classes, class distributions, and feature dimensions before and after preprocessing. 
All datasets contain matched patient-level measurements across three omics modalities: DNA methylation, mRNA expression, and miRNA expression. Only samples with complete measurements in all modalities are retained for analysis. 
The datasets display varying degrees of class imbalance, including severely underrepresented subtypes such as HER2-enriched in BRCA and MSEAC in UCEC, offering a realistic and challenging evaluation setting.
Although all datasets originate from TCGA, they represent biologically distinct cancer types with different imbalance structures and molecular characteristics. Figure~\ref{fig:omicsHistogram} shows histograms for DNA methylation, mRNA, and miRNA features, illustrating substantial heterogeneity across datasets and modalities. 
These datasets also reflect the typical small-$n$, large-$p$ regime of multi-omics studies, where a few hundred patients are measured on tens of thousands of features, making robust graph learning, fusion, and regularization essential.

\begin{table}[h]
\caption{Hyperparameter tuning ranges for TF-DWGNet on the BRCA, UCEC, and KIPAN datasets. The optimal choices selected for the final model are shown in bold.}
\label{tab:hyperparams}
\centering
\begin{tabular}{lr}
\toprule
\textbf{Hyperparameter} & \textbf{Tuning Range} \\
\midrule
\multicolumn{2}{l}{\textit{XGBoost Parameters}} \\
Number of Trees ($M$) & \textbf{100}, 200, 500, 1000 \\
\midrule
\multicolumn{2}{l}{\textit{Neural Network Architecture}} \\
Depth (hidden layers) & 1, \textbf{2}, 3 \\
Width (neurons per layer) & 32, \textbf{64}, 128 \\
Activation Function & \textbf{ReLU}, leakyReLU \\
\midrule
\multicolumn{2}{l}{\textit{Training Parameters}} \\
Learning Rate & 0.001, 0.0005, \textbf{0.0001} \\
Batch Size & 8, 16, 32, \textbf{64} \\
Training Epochs & 500, \textbf{1000}, 1500 \\
\midrule
\multicolumn{2}{l}{\textit{Regularization and Early Stopping}} \\
Dropout Rate & 0.2, 0.3, \textbf{0.5} \\
L2 Regularization Lambda & \textbf{0.01} \\
Early Stopping Patience & 5, \textbf{10} \\
Early Stopping Min Delta & \textbf{0.001}, 0.005 \\
\midrule
\multicolumn{2}{l}{\textit{Low-Rank Decomposition}} \\
Rank ($R$) & 4, 8, 16, 24, 30, 32, 40, \textbf{48}, 50, 64 \\
\midrule
\multicolumn{2}{l}{\textit{Deep Residual Network}} \\
Number of Residuals & 1, 2, \textbf{3}, 5 \\
\bottomrule
\end{tabular}
\end{table}

\begin{table}[h]
\caption{Reduced features (nodes), edge counts ($E_i$), and edge-to-node ratio ($m_i$) for graphs constructed by TF-DMGNet on BRCA, UCEC, and KIPAN. For each omics type $i \in \{1,2,3\}$, $p_i^*$ denotes the number of reduced features, $|E_i|$ the number of edges, and $m_i = |E_i| / p_i^*$ the edge-to-node ratio.
}
\label{tab:graphstats}
\centering
\setlength{\tabcolsep}{1mm}
\begin{tabular}{@{}lrrr@{}}
\toprule
\textbf{Dataset} & \multicolumn{1}{c}{\textbf{Nodes}} & \multicolumn{1}{c}{\textbf{Edges}} & \multicolumn{1}{c}{\textbf{Edge/Node}} \\
& ($p_1^*$:$p_2^*$:$p_3^*$) & ($|E_1|$:$|E_2|$:$|E_3|$) & ($m_1$:$m_2$:$m_3$) \\
\midrule
BRCA & 878:825:494 & 4300:3658:4176 & 4.90:4.43:8.45\\
UCEC & 381:336:272 & 882:763:794  & 2.31:2.27:2.92\\
KIPAN & 311:244:155 & 749:552:437 & 2.41:2.26:2.82\\
\bottomrule
\end{tabular}
\end{table}

\begin{table*}[h!] 
\caption{Classification performance comparison on BRCA, UCEC, and KIPAN. Metrics are reported as mean ± standard deviation [95\% confidence interval] over 20 reproducible seeds. Best results are bolded. Superscripts $^{*}$, $^{**}$, and $^{***}$ indicate that TF-DWGNet is statistically significantly better than the corresponding model (p-value $<$ 0.05, $<$ 0.01, and $<$ 0.001, respectively) based on Welch's $t$-test. Note that KIPAN is included as a simpler, proof-of-concept dataset due to its relatively easy classification structure for all models.}
\label{tab:classification}
\centering
\setlength{\tabcolsep}{1mm}
\begin{tabular}{@{}lllll@{}}
\toprule
\textbf{Dataset} & \textbf{Model} & \textbf{Accuracy} & \textbf{F1-weighted} & \textbf{F1-macro} \\
\midrule
\multirow{5}{*}{BRCA} 
& RF & 0.747 ± 0.026 [0.735, 0.759] $^{***}$ & 0.736 ± 0.029 [0.722, 0.750] $^{***}$ & 0.671 ± 0.036 [0.654, 0.688] $^{***}$\\
& XGBoost & 0.742 ± 0.021 [0.732, 0.752] $^{***}$ & 0.731 ± 0.024 [0.720, 0.742] $^{***}$ & 0.664 ± 0.041 [0.645, 0.683] $^{***}$\\
& DFN & 0.751 ± 0.021 [0.741, 0.761] $^{***}$ & 0.750 ± 0.021 [0.740, 0.760] $^{***}$ & 0.705 ± 0.033 [0.689, 0.721] $^{***}$\\
& GCN & 0.752 ± 0.028 [0.739, 0.765] $^{***}$ & 0.751 ± 0.029 [0.737, 0.765] $^{***}$ & 0.708 ± 0.043 [0.688, 0.728] $^{***}$\\
& GEDFN & 0.790 ± 0.025 [0.778, 0.801] $^{***}$ & 0.792 ± 0.024 [0.781, 0.804] $^{***}$ & 0.748 ± 0.032 [0.733, 0.763] $^{**}$ \\
& $\mathrm{TF\text{-}DWGNet}_{\text{rf}}$ & 0.789 ± 0.033 [0.774, 0.804] $^{**}$ & 0.791 ± 0.032 [0.776, 0.806] $^{***}$ & 0.748 ± 0.038 [0.730, 0.766] $^{**}$ \\
& $\mathrm{TF\text{-}DWGNet}_{\text{con}}$ & 0.806 ± 0.031 [0.791, 0.821] & 0.814 ± 0.030 [0.800, 0.828] 
& 0.782 ± 0.040 [0.763, 0.801] \\
& $\mathrm{TF\text{-}DWGNet}_{\text{undir}}$ & 0.800 ± 0.030 [0.786, 0.814] $^{*}$ & 0.804 ± 0.029 [0.790, 0.818] $^{*}$ & 0.765 ± 0.039 [0.747, 0.783] \\
& TF-DWGNet & \textbf{0.821 ± 0.023 [0.810, 0.832]} & \textbf{0.826 ± 0.022 [0.816, 0.836]} & \textbf{0.785 ± 0.033 [0.769, 0.801]} \\
\midrule
\multirow{5}{*}{UCEC}
& RF & 0.852 ± 0.022 [0.842, 0.862] $^{*}$ & 0.824 ± 0.025 [0.812, 0.836] $^{***}$ & 0.544 ± 0.024 [0.533, 0.555] $^{***}$ \\
& XGBoost & 0.860 ± 0.026 [0.848, 0.872] & 0.839 ± 0.026 [0.827, 0.851] $^{*}$ & 0.559 ± 0.022 [0.549, 0.569] $^{***}$\\
& DFN & 0.842 ± 0.034 [0.826, 0.858] $^{**}$ & 0.837 ± 0.031 [0.822, 0.852] $^{*}$ & 0.566 ± 0.026 [0.554, 0.578] $^{**}$\\
& GCN & 0.837 ± 0.038 [0.819, 0.855] $^{**}$ & 0.833 ± 0.035 [0.817, 0.849] $^{**}$ & 0.563 ± 0.033 [0.548, 0.578] $^{**}$\\
& GEDFN & 0.859 ± 0.024 [0.848, 0.871] & 0.848 ± 0.022 [0.838, 0.858] & 0.571 ± 0.018 [0.562, 0.579] $^{*}$ \\
& $\mathrm{TF\text{-}DWGNet}_{\text{rf}}$ & 0.859 ± 0.031 [0.844, 0.874] & 0.846 ± 0.027 [0.833, 0.859] & 0.574 ± 0.020 [0.565, 0.583] \\
& $\mathrm{TF\text{-}DWGNet}_{\text{con}}$ & 0.864 ± 0.031 [0.849, 0.879] & 0.853 ± 0.030 [0.839, 0.867] & 0.587 ± 0.045 [0.566, 0.608] \\
& $\mathrm{TF\text{-}DWGNet}_{\text{undir}}$ & 0.866 ± 0.031 [0.851, 0.881] & 0.853 ± 0.027 [0.840, 0.866] & 0.578 ± 0.023 [0.567, 0.589] \\
& TF-DWGNet & \textbf{0.869 ± 0.038 [0.851, 0.887]} & \textbf{0.858 ± 0.034 [0.842, 0.874]} & \textbf{0.598 ± 0.056 [0.572, 0.624]} \\
\midrule
\multirow{5}{*}{KIPAN}
& RF & 0.953 ± 0.023 [0.942, 0.964] & 0.953 ± 0.023 [0.942, 0.964] & 0.943 ± 0.033 [0.928, 0.958] \\
& XGBoost & 0.952 ± 0.023 [0.941, 0.963] $^{*}$ & 0.952 ± 0.023 [0.941, 0.963] $^{*}$ & 0.942 ± 0.034 [0.926, 0.958] \\
& DFN & 0.957 ± 0.017 [0.949, 0.965] & 0.957 ± 0.017 [0.949, 0.965] & 0.945 ± 0.024 [0.934, 0.956] \\
& GCN & 0.955 ± 0.023 [0.944, 0.966] & 0.955 ± 0.023 [0.944, 0.966] & 0.945 ± 0.030 [0.931, 0.959] \\
& GEDFN & 0.957 ± 0.017 [0.949, 0.965] & 0.957 ± 0.017 [0.949, 0.965] & 0.946 ± 0.025 [0.934, 0.957] \\
& $\mathrm{TF\text{-}DWGNet}_{\text{rf}}$ & 0.957 ± 0.020 [0.948, 0.966] & 0.957 ± 0.020 [0.948, 0.966] & 0.945 ± 0.024 [0.934, 0.956] \\
& $\mathrm{TF\text{-}DWGNet}_{\text{con}}$ & 0.960 ± 0.016 [0.953, 0.967] & 0.960 ± 0.016 [0.953, 0.967] & 0.947 ± 0.024 [0.936, 0.958] \\
& $\mathrm{TF\text{-}DWGNet}_{\text{undir}}$ & 0.958 ± 0.020 [0.949, 0.967] & 0.959 ± 0.019 [0.950, 0.968] & 0.947 ± 0.026 [0.935, 0.959] \\
& TF-DWGNet & \textbf{0.964 ± 0.017 [0.956, 0.972]} & \textbf{0.964 ± 0.017 [0.956, 0.972]} & \textbf{0.956 ± 0.024 [0.945, 0.967]} \\
\bottomrule
\end{tabular}
\end{table*}

\begin{figure*}[h]
  \centering
  \includegraphics[width=0.9\textwidth]{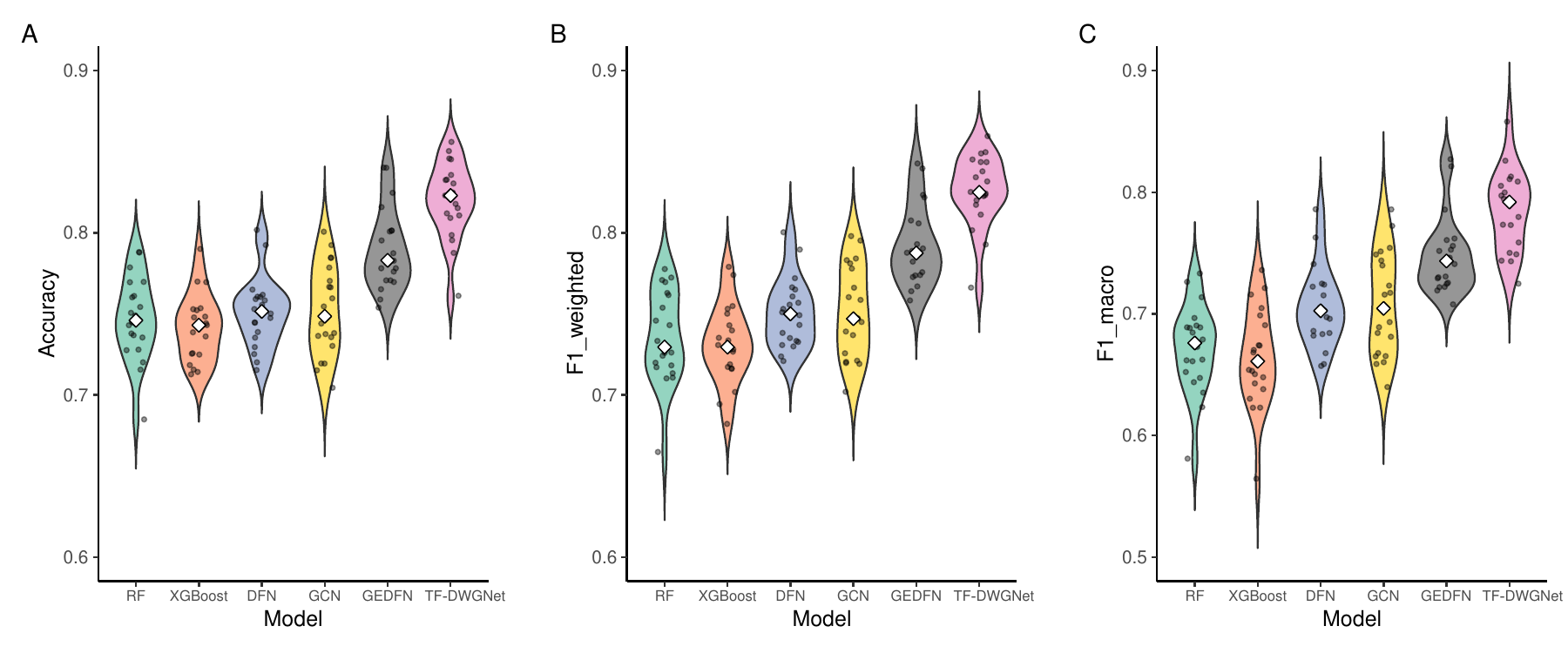}
  \caption{Performance comparison of TF-DWGNet and baseline models across 20 independent seeds on the BRCA dataset using violin plots. Each violin displays score distributions, individual values (dots), and medians (white diamonds). Results are shown for (A) Accuracy, (B) F1-weighted, and (C) F1-macro.} \label{fig:BRCAViolinPlot}
\end{figure*}

\noindent \textbf{Baselines, Ablation Studies, and Experimental Setting.}
We compare TG-DWGNet to several conventional and deep learning baselines, including RF, XGBoost, DFN, GCN, and GEDFN, as well as \textit{\textbf{three ablation variants}} designed to isolate the effects of graph construction, fusion strategy, and directionality: 
(1) $\boldsymbol{\mathrm{TF\text{-}DWGNet}_{\text{rf}}}$ replaces XGBoost with RF for graph construction, examining the effect of the tree-based estimator; (2) $\boldsymbol{\mathrm{TF\text{-}DWGNet}_{\text{con}}}$ replaces tensor fusion with direct concatenation, assessing the added value of tensor modeling and CP decomposition; and (3) $\boldsymbol{\mathrm{TF\text{-}DWGNet}_{\text{undir}}}$ symmetrizes the adjacency matrix to remove directional information, evaluating the importance of task-conditioned directionality. 
All baselines are trained on column-wise concatenated multi-omics features, unless the model inherently uses graph structure. For fairness, both the GCN and GEDFN use XGBoost-derived graphs as TF-DWGNet. All models are evaluated using the same 20 stratified training, validation, and test splits (60\%/20\%/20\%) with fixed random seeds. 
For tree-based baselines, XGBoost and RF use 100 estimators. DFN, GCN and GEDFN use two hidden layers of width 64. 
TG-DWGNet employs one hidden layer of width 64 in each GNN branch and three residual blocks in the DFN, each with three hidden layers of width 64. Training is performed using a learning rate of 0.0001, a batch size of 64, and a maximum of 1000 training epochs. Regularization strategies include dropout (0.5), L2 regularization ($\lambda = 0.01$), and early stopping (patience = 10 epochs, minimum validation-loss improvement = 0.001). The tensor fusion module uses a CP rank of $R = 48$. 
The default TF-DWGNet model uses 100 trees in the XGBoost graph-construction module, which performed reliably in preliminary experiments. Table~\ref{tab:hyperparams} summarizes the hyperparameter search space and the selected optimal values.

\noindent \textbf{Evaluation Metrics.} 
We evaluate the model performance using Accuracy, F1-weighted, and F1-macro scores.
F1-macro computes the F1-score for each class independently and averages them equally, providing a balanced view of performance across all classes. In contrast, F1-weighted computes a weighted average of per-class F1-scores, with weights proportional to class frequencies, making it more sensitive to class imbalance. For a classification problem with $C$ classes, the generalized F1-scores are computed as:
$$F1 = \sum_{c=1}^C \left( w_c \times \frac{2 \times \text{precision}_c \times \text{recall}_c}{\text{precision}_c + \text{recall}_c} \right),$$
where 
$$\text{precision}_c = \frac{TP_c}{TP_c + FP_c}, \text{recall}_c = \frac{TP_c}{TP_c + FN_c}.$$ 
The class weights $w_c$ follow $w_c =\frac{1}{C}$ for F1-macro, and $w_c=\frac{n_c}{n}$ for F1-weighted, with $n_c$ the number of samples in class $c$. 
We report the mean, standard deviation (SD), and 95\% confidence intervals (CIs) over 20 independent training-validation-test splits. To assess statistical significance, we also perform pairwise Welch's $t$-tests between TF-DWGNet and each comparator model, which account for unequal variances across runs.

\subsection{Results and Discussion}

\textbf{Constructed Graph Properties.}
Table~\ref{tab:graphstats} summarizes the reduced feature dimensions $p_i^*$ for each omics modality ($i = 1, 2, 3$), which correspond to the number of nodes $|V_i|$ in the constructed directed weighted graphs. The resulting feature sets range from roughly 150 to 900 nodes, reflecting adaptive compression across datasets. The table also reports edge counts $|E_i|$. The learned graphs remain sparse, with edge-to-node ratios $m_i = |E_i|/{p_i^*}$ between two and nine, yet are sufficiently connected to preserve modality-specific structure. These properties enable effective message passing in downstream GNNs while maintaining computational efficiency.

\begin{figure}[h]
  \centering
\includegraphics[width=\columnwidth]{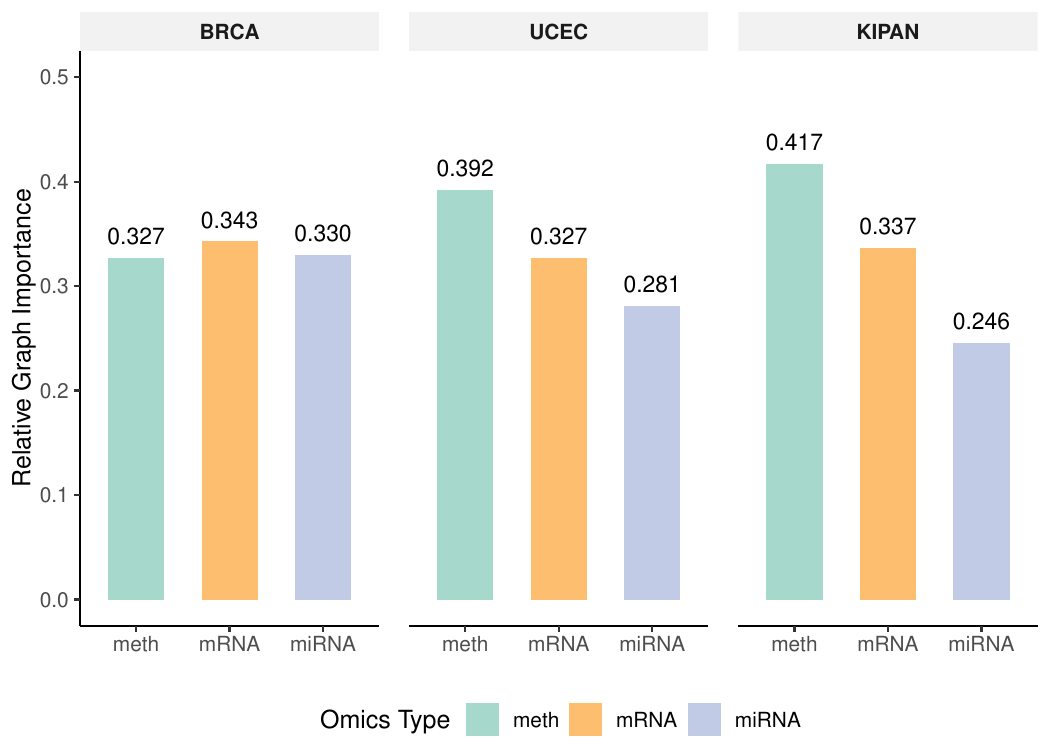}
  \caption{Relative importance of different omics types (DNA methylation, mRNA, and miRNA) in TF-DWGNet's classification performance across BRCA, UCEC, and KIPAN. 
  Each bar reflects the normalized contribution of an omics graph, automatically learned by the model. 
  } \label{fig:graphImportance}
\end{figure}

\noindent \textbf{Classification Performance.} 
Table~\ref{tab:classification} presents the classification performance of TF-DWGNet, its three ablation variants, and all baselines on BRCA, UCEC, and KIPAN. We report Accuracy, F1-weighted, and F1-macro scores, with the best results shown in bold. Overall, TF-DWGNet consistently achieves the strongest performance across all datasets and metrics, demonstrating robustness and predictive strength in heterogeneous multi-omics settings.
On BRCA, TF-DWGNet achieves substantial gains, including more than an 8\% increase in accuracy relative to the weakest baseline and even larger improvements in F1-macro and F1-weighted (approximately 12\% and 10\%, respectively). These gains highlight the model's ability to handle subtype imbalance and capture both intra- and inter-modal structure. All ablation variants underperform the full model, though to varying degrees: $\mathrm{TF\text{-}DWGNet}_{\text{rf}}$ shows the largest degradation, reflecting the importance of XGBoost-based graph construction; $\mathrm{TF\text{-}DWGNet}_{\text{undir}}$ performs significantly worse on accuracy and F1-weighted, underscoring the benefit of modeling directionality; $\mathrm{TF\text{-}DWGNet}_{\text{con}}$ trails by a smaller margin, but TF-DWGNet still attains the highest means, suggesting additional benefit from tension-based feature integration.
On UCEC and KIPAN, TF-DWGNet again achieves the best performance across all metrics, with consistent but smaller margins of 1-3\%. The ablation variants remain competitive but uniformly below the full model, confirming that each architectural component contributes to predictive performance.
To assess statistical significance, we perform pairwise Welch's $t$-tests based on 20 independent runs per model, comparing TF-DWGNet with each baseline and ablation. Superscripts in Table~\ref{tab:classification} indicate the corresponding significance levels. For BRCA, nearly all improvements are highly significant (p-value $<$ 0.001), and the confidence intervals for TF-DWGNet lie distinctly above competing models. Two ablation variants ($\mathrm{TF\text{-}DWGNet}_{\text{rf}}$ and $\mathrm{TF\text{-}DWGNet}_{\text{undir}}$) also differ significantly from TF-DWGNet, highlighting the added value of directional information and XGBoost-based graph construction.
On UCEC, several differences with classical baselines remain significant, while none of the ablation variants differ significantly from the full model.
For KIPAN, a few differences are significant, consistent with all models achieving strong performance on this simpler task.

Figure~\ref{fig:BRCAViolinPlot} visualizes the distribution of TF-DWGNet and the baseline performance across 20 BRCA splits. 
TF-DWGNet achieves the highest median for all metrics and exhibits tight distributions centered near the median, indicating high stability. Most of the TF-DWGNet's individual run scores exceed the maximum values achieved by other baselines, demonstrating that it outperforms the baselines not only on average but also on nearly every individual test split. 

\noindent \textbf{Feature Interpretability.}  
Each input feature (biomarker) receives a ranking score after training. While the absolute magnitude of the score is not directly interpretable, the relative ranking provides meaningful insight into each feature's contribution to subtype discrimination. High-ranking biomarkers may indicate strong statistical associations or potential mechanistic links to specific cancer types. 
For BRCA, several top-ranked features identified by TF-DWGNet align with established cancer biology. For example, \textit{FAM134B}, an ER-phagy receptor associated with hypoxic breast cancer survival, has been shown to suppress tumor proliferation when silenced \cite{chipurupalli2022cancer}. Likewise, \textit{FOXC1}, commonly overexpressed in basal-like breast cancer (e.g., BRCA1-mutant tumors), promotes proliferation and increases sensitivity to PARP inhibitors \cite{johnson2016foxc1}. Such concordance supports TF-DWGNet's ability to highlight biologically meaningful, subtype-relevant markers. 
Although our interpretability validation focuses on agreement with known biomarkers, a more systematic analysis, such as pathway enrichment or comparisons with curated biological knowledge, remains outside the scope of this study. TF-DWGNet is designed to provide statistically grounded, biologically plausible signals that can guide downstream investigations. In practice, groups of highly ranked biomarkers may reveal coordinated pathway activity or shared regulatory mechanisms that warrant further follow-up. 

\noindent \textbf{Modality Interpretability.} Figure~\ref{fig:graphImportance} shows the relative contribution of each omics modality to TF-DWGNet predictions, derived from the aggregated feature-importance scores. The model adaptively reweights each modality based on its predictive utility. In BRCA, DNA methylation, mRNA expression, and miRNA expression contribute at comparable levels, suggesting complementary and non-redundant information across modalities.  In contrast, for UCEC and KIPAN, DNA methylation emerges as the dominant source of predictive signal, while miRNA plays a comparatively smaller role. 
These patterns demonstrate TF-DWGNet's ability to integrate heterogeneous omic sources in a data-driven manner,  allowing the model to emphasize biologically relevant modalities without manual weighting or hand-engineered assumptions. While our analysis emphasizes statistical interpretability, validating modality-level contributions against known molecular mechanisms remains an important direction for future work. The supervised graph construction mechanism is designed to capture task-relevant dependencies, providing a principled basis for the learned modality contributions and directional interactions.

\section{Conclusion}
We propose TF-DWGNet, a novel graph neural network framework for multi-omics cancer subtype classification that integrates directed weighted graph learning with tensor-based feature fusion in an end-to-end supervised architecture. By jointly modeling modality-specific graph structures and higher-order cross-modal interactions, TF-DWGNet provides a unified and expressive approach to capture the complexities inherent in heterogeneous omics data. The framework is also modality-flexible, enabling the incorporation of additional omics types without modifying its core design and supporting future extensions to broader multi-omics settings.
Comprehensive experiments on three real-world cancer datasets demonstrate that TF-DWGNet consistently outperforms strong classical and deep-learning baselines across all major evaluation metrics. These results highlight the model's robustness, predictive accuracy, and its ability to generalize across cancers with distinct molecular architectures and varying degrees of class imbalances. 
Beyond improved predictive performance, TF-DWGNet provides interpretable outputs, including modality-level contribution scores and biomarker-level importance rankings. These insights help reveal potential molecular drivers and dataset-specific patterns, offering interpretability essential for downstream biological investigations and clinical hypothesis generation. Overall, TF-DWGNet offers a powerful and interpretable solution for multi-omics integration, with significant potential to support precision oncology and deepen our understanding of cancer subtype heterogeneity.

\section*{Data and Code Availability}
The original datasets are from TCGA Broad GDAC Firehose (\url{https://gdac.broadinstitute.org/}). The harmonized datasets used in this study (BRCA, UCEC, and KIPAN) and the code of our model (TF-DWGNet) are available upon publication.

\begin{acks}
This work is partially supported by the National Institute of General Medical Sciences of the National Institutes of Health (NIH/NIGMS) under Award P20GM104420. 
\end{acks}


\bibliographystyle{main}
\bibliography{main}

\end{document}